\documentclass[prd,twocolumn,showpacs,superscriptaddress]{revtex4-1}
\usepackage{lipsum}
\usepackage{hyperref}
\usepackage{graphicx}
\usepackage{graphics}
\usepackage{amsmath}
\usepackage{amssymb}
\usepackage{amsthm}
\usepackage{mathbbol}
\usepackage{amsfonts}

\newcommand{\be}{\begin{equation}}
\newcommand{\ee}{\end{equation}}
\newcommand{\bea}{\begin{eqnarray}}
\newcommand{\eea}{\end{eqnarray}}
 
\begin{document}

\title{A novel analysis method for excited states in lattice QCD - the nucleon case}

\author{C. Alexandrou}
\affiliation{Department of Physics, University of Cyprus, P.O. Box 20537, 1678 Nicosia, Cyprus}
\affiliation{The Cyprus Institute, P.O. Box 27456, 1645 Nicosia, Cyprus}

\author{T. Leontiou}
\affiliation{General Department, Frederick University, 1036 Nicosia, Cyprus}

\author{C. N. Papanicolas}
\affiliation{The Cyprus Institute, P.O. Box 27456, 1645 Nicosia, Cyprus}
\affiliation{Department of Physics and Institute of Accelerating Systems Applications, University of Athens, Athens, Greece}

\author{E. Stiliaris}
\affiliation{Department of Physics and Institute of Accelerating Systems Applications, University of Athens, Athens, Greece}

\begin{abstract}
We employ a novel method to analyze Euclidean correlation functions
entering the calculation of hadron energies in lattice QCD. The method
is based on the sampling of all possible solutions allowed by the
spectral decomposition of the hadron correlators.  We demonstrate the
applicability of the method by studying the nucleon excited states in
the positive and negative parity channels over a pion mass range of
about 400~MeV to 150~MeV.  The results are compared to the standard
variational approach routinely used to study excited states within
lattice QCD. The main advantage of our new approach is its ability to
unambiguously determine all excited states for which the Euclidean
time correlation function is sensitive on.
\end{abstract}

\maketitle

\section{Introduction}

The study of excited states within the framework of Quantum
Chromodynamics on the lattice (LQCD) is difficult since it is based on
the evaluation of Euclidean correlation functions for which the
excited states are exponentially suppressed as compared to the ground
state. The standard approach to study excited states is based on the
variational principle, where one considers a number of interpolating
fields as a variational basis and defines a generalized eigenvalue
problem (GEVP) using the correlation matrix computed within the chosen
variational basis.  The GEVP has been widely applied in the study of
hadron spectroscopy by a number of lattice groups with recent results
given in
Refs.~\cite{Basak:2007kj,Gattringer:2008be,Gattringer:2008vj,Mahbub:2010jz,Bulava:2010yg,Alexandrou:2013fsu}.

In this paper, we examine in depth the application of a new method
based on statistical concepts for extracting the excited states from
Euclidean correlators to study the nucleon spectrum.  The so called
Athens Model Independent Analysis Scheme (AMIAS)
~\cite{Papanicolas:2012sb} was originally developed to extract
scattering amplitudes from experimental measurements in the $N$ to
$\Delta$ transition~\cite{Papanicolas:2007} without utilizing a specific model thus avoiding
model dependencies of the outcome.  A first analysis of the nucleon
two-point function in LQCD was carried out with promising results in
Ref.~\cite{Alexandrou:2008bp}.  In this work, we extend the method to
analyze correlation matrices and compare the results with those
obtained in our recent study using the standard variational
method~\cite{Alexandrou:2013fsu}. An advantage of AMIAS is that it can
be applied to the correlation function at small separation times by
allowing any number of states to contribute rather than to the large
time-limit behavior typically done in the variational method. In fact,
the merit of the method is that it determines the actual number of
states on which the correlation matrix is sensitive on.  Thus AMIAS
does not rely on plateau identification of effective masses, which are
usually noisy and thus difficult to determine, but instead it utilizes
all the information encoded in the correlation function with the
advantage of exploiting the small time separations where the
statistical errors are small.

The paper is organized as follows: In section~\ref{sec:method} we give
the general description of the method, including a discussion of an
importance sampling version of the algorithm. The method is compared
to standard variational approach paying particular attention to the
determination of the number and correlations of the fitting parameter,
which are advantages of our method. Section
\ref{sec:lattice_techniques} gives the lattice QCD formulation
pertinent to the problem at hand.  Finally, in
section~\ref{sec:results}, we discuss the results extracted using
AMIAS and in section~\ref{sec:conclusions} we summarize our findings
and give our conclusions.

\section{The AMIAS approach}
\label{sec:method}
  In this section we introduce AMIAS and demonstrate its applicability
  in extracting the excited states from Euclidean correlation
  functions.

The starting point of AMIAS is a set of measurements with the
associated statistical errors~\cite{Papanicolas:2012sb}.  Since the quantity of interest in the
work is a Euclidean two-point function computed at discrete times we
explain the method based on the measurement of these correlators.  Let
us denote by $C^k(t_j), j=1,\cdots,N_t$ the $k^{\rm th}$ measurement
at time $t_j$.  The next ingredient for AMIAS is a theoretical
parameterization of the measurements or an 'underlining theory'. For
lattice QCD correlators this is well-known from the spectral
decomposition of the hadron propagator.  
\be
\bar{C}(t_j)=\sum_{n=0}^\infty A_n e^{-E_n t_j}\quad.
\label{correlator0}
\ee
The exact correlator is approximated as the average over $N$ measurements of
an appropriately defined quantity to be described in Section~\ref{sec:lattice_techniques}: 
\be
 {C}(t_j)=\sum_{k=1}^N C^k(t_j)
\ee
for a total of $N$ measurements.
The parameters $A_n$ and $E_n$ are to be determined by AMIAS, where we order the exponentials by the value of $E_n$
$E_0< E_1< E_2,\cdots$. For large values of the time the exponential with the smallest exponent dominates.

More details
on the lattice formulation will be given in Section~\ref{sec:lattice_techniques}.

\begin{figure}[ht]
\includegraphics[width=3in]{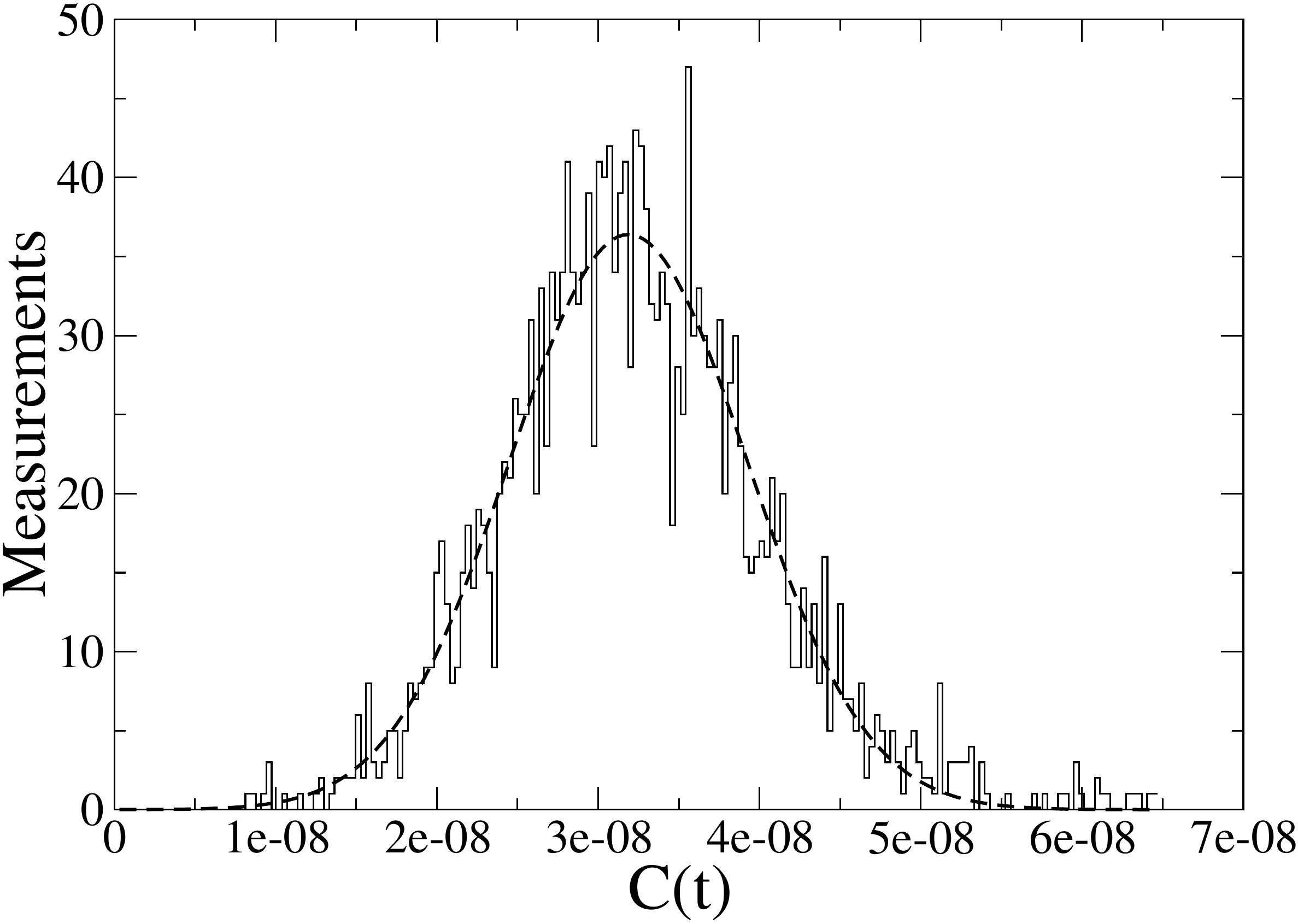}
\caption{
The correlator at a given $t_j$ is computed using   5000 measurements, having an average value of $C(t_j)=3.2\times 10^{-8}$.
The standard jackknife error is $1.8\times 10^{-10}$ and the weighted standard deviation is $\sigma_j/\sqrt{N}=1.7\times 10^{-10}$,
thus demonstrating that the measurements are uncorrelated. 
\label{fig1}
}
\end{figure}

 The Central Limit Theorem states that the probability for the above
 average at time $t$ to have a value equal to $\bar{C}(t)$ is
\be P(\bar{C}(t_j))=B_j \exp{\left(-\frac{(\bar{C}(t_j)-
    C(t_j))^2}{2(\sigma_j/\sqrt{N})^2}\right)},
\label{central}
\ee 
 where $\sigma_j$ is the standard
deviation at time $t_j$ and $B_j$ is, for our purpose, an irrelevant normalization constant.  The above result can be found in many textbooks and the 
assumptions made are that the measurements are uncorrelated and that
$N$ is large. The value of $\sigma_j$ in Eq.~(\ref{central}) will,
in general, be underestimated if correlations are present. For LQCD 
the correct value of $\sigma_j$ is usually obtained using either a jackknife
or bootstrap procedure.  An example of the distribution given in Eq.~(\ref{central}) 
for  a given value of $t_j$
is shown in Fig.~\ref{fig1} showing indeed that the 
distribution is to a good accuracy a Gaussian. The 
jackknife error is found to be in agreement with that predicted by the
Central Limit Theorem, which means that the measurements used for
the average are uncorrelated.

The probability of obtaining a set of $N_t$ measurements,
each for a different value of $t_j$, is equal to the product of the
probabilities for each data point: 
\be
\begin{split}
&P(C(t_j); \forall
j=1,N_t)=\prod_{k=1}^{N_t}
P(C(t_k))\\
&=\exp{\left(-\sum_{k=1}^{N_t}\frac{(\bar{C}(t_k)-
    C(t_k))^2}{2(\sigma_k/\sqrt{N})^2}\right)}\,, \end{split} \ee
where we have set $B_j=1$.
 If we use
the expansion in terms of exponentials given in Eq.~(\ref{correlator0}), then the
above result can be written as 
\bea \label{chi2} P({C}(t_j);
\forall
j&=&1,N_t)=e^{-\frac{\chi^2}{2}},\\
~~{\rm where}~~~\chi^2&=&\sum_{k=1}^{N_t}\frac{({C}(t_k)-\sum_{n=0}^{\infty} A_n
  e^{-E_n t_k})^2}{(\sigma_k/\sqrt{N})^2}. \nonumber
\eea 
For practical purposes the infinite summation over the energy spectrum
involving the parameters $A_n$ and $E_n$ must be truncated keeping a 
 maximum number of $n_{\rm max}$  energy levels:
\be \label{tildechi2} 
\chi^2\sim\tilde{\chi}^2(n_{\rm max})=\sum_{k=1}^{N_t}\frac{(C(t_k)-\sum_{n=0}^{n_{\rm max}} A_n
  e^{-E_n t_k})^2}{(\sigma_{k}/\sqrt{N})^2}, 
\ee
The $\tilde{\chi}^2$
appearing in the above equation is the usual $\chi^2$ used in the
least-squares fitting  method.  The Probability Density Function (PDF) of Eq.~(\ref{chi2}) is the basis of the
$\chi^{2}$ minimization in the case where the values of the parameters
are unknown: Maximizing the probability is equivalent to minimizing
the exponent. 

The principal idea behind AMIAS ~\cite{Papanicolas:2012sb} is that
given a particular set of sampled averages, $C(t)$, 
any arbitrary value assigned to the parameters $A_n$ and $E_n$ 
constitutes a solution of Eq.~\ref{correlator0} having the probability 
given by Eq.~(\ref{chi2}). Following standard statistical
concepts~\cite{Papanicolas:2012sb}, the probability $\Pi(a_i)$ that
the parameter $A_i$ assumes a specific value $a_i$ in the range
$(b_i,c_i)$ is equal to

\be \label{critical} \Pi(a_i)=\frac{
  \int_{b_i}^{c_i}\, dA_i\,
  \int_{-\infty}^\infty\ dE_i\,
  \int_{-\infty}^\infty\,\left(\prod_{\substack{j=1\\j\ne i}}^n dE_j
  dA_j\right) \, A_i e^{-\tilde{\chi}^2/2}}
{\int_{-\infty}^\infty\,\left(\prod_{j=1}^n dA_j \,dE_j \right)\,A_i
  e^{-\tilde{\chi}^2/2}}.  
\ee

The above formulation has been applied to the analysis of experimental
nuclear extrapolation data~\cite{Papanicolas:2007} and to a lattice QCD 
calculation of the nucleon excited states~\cite{Alexandrou:2008bp}, 
using a  uniform sampling for values of $A_i\in [b_i,c_i)$ in Eq.~(\ref{critical}), yielding consistent results with the standard analysis.

For reasons that  will become apparent later on, instead of using a single correlation function
we can apply AMIAS to 
a correlation matrix with the spectral decomposition
\be
\label{correlator}
\bar{C}^{(i,j)}(t)=\sum_{n=0}^\infty A_n^{(i)}(A_n^{(j)})^\dagger e^{-E_n t}\,,\,\,\,
  i,j=1,\ldots, N_{\rm inter}. \nonumber
\ee
As in the case of a single correlator we  keep a maximum number of $n_{\rm max}$
energy levels in the sum.

The method can be applied to a general matrix and in the analysis
where we apply AMIAS we will be interested in correlation matrices
that are hermitian. This constrains the number of amplitudes $A^{(i)}$, 
simplifying the fitting problem.
Although the matrix elements
can be correlated with each other 
Eq.~(\ref{tildechi2}) is valid for each matrix element independently,
hence the PDF for the
entire correlation matrix can be expressed as the product of the PDF for each
matrix-element, leading to a $\chi^2$ that is the sum of the
individual $\chi^2$'s: 

\be \label{chi2ij}
\begin{split}
&P(C^{(i,j)}(t_n); \forall i,j,n)=e^{-\frac{\chi^2}{2}},\\&{\rm where}~~~\\ 
&\chi^2=\sum_{i,j}\sum_{k=1}^{N_t}\frac{(C^{(i,j)}(t_k)-\sum_{n=0}^{\infty}
  A^{(i)}_n(A^{(j)}_n)^\dagger
  e^{-E_nt_k})^2}{(\sigma_{t_k}^{(i,j)}/\sqrt{N})^2}. 
\end{split}
\ee

\subsection{Importance sampling}
While uniform sampling works well for a single correlator, for a correlation
matrix 
determining the fit parameters $A^{(j)}$ and $E^{(j)}$ is impractical  due to the large number of parameters. One can use the
Metropolis algorithm~\cite{metropolis} to sample more efficiently the distribution
$e^{-\tilde{\chi}/2}$, through a random walk.
However, the Metropolis algorithm
being a local random walk has the disadvantage of getting trapped at
local minima. In order to avoid this problem and ensure access to the
full multidimensional space, we employ parallel
tempering~\cite{tempering}.

In our parallel tempering scheme we run a number $r$ of additional
random walks, randomly initialized at different `temperatures'. Thus we have a chain of
random walks sampling the sequence of PDFs: \newline 
\(\{\exp(-\frac{\chi^2}{2}),\exp(-\frac{\chi^2}{2T_1}),\exp(-\frac{\chi^2}{2T_2}),...,\exp(-\frac{\chi^2}{2T_r})\}\), 
where the $T_i$s are the parallel tempering temperatures. Sampling the
distribution histogram of each parameter is done through
Eq.~\ref{critical} in the usual way but only the original PDF is finally used
obtained  by setting $T_i=1$.  The high temperature PDFs are generally able to
sample larger phase space, whereas the exact PDF with $T=1$
whilst having precise sampling in a local region of phase space, may
become trapped in local minima. Information from the high temperature walks is passed down to the low
temperature ones through exchanges among ensembles. Swaps are
normally attempted between systems with adjacent temperatures, $k=i$ and
$k=i+1$, and are accepted with  probability 
\be
\min\{1,  \exp{ \left(\left( \frac{1}{T_i} - \frac{1}{T_{i+1}} \right) \left( \chi^2(i)/2-\chi^2(i+1)/2\right) \right) }. \}
\ee
Parallel tempering is an exact method, in that it satisfies the
detailed balance condition.

\subsection{Number of parameters} 

\begin{figure}[ht]
  \medskip
    \includegraphics[width=3in, keepaspectratio]{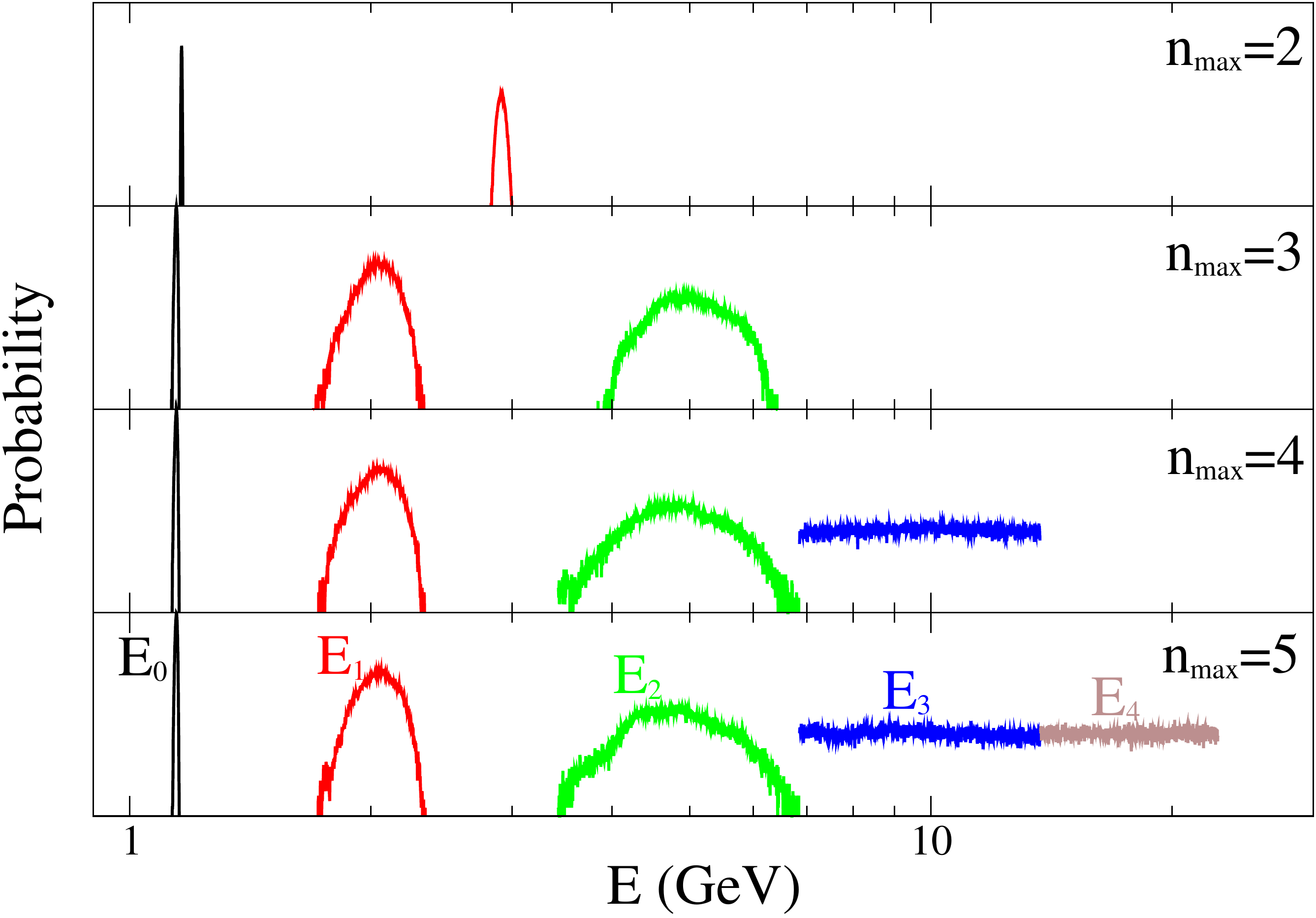}
    \caption{\label{nparams} The Probability Density Functions (PDFs)
      of $E_n$ for different values of the truncation parameter
      $n_{\rm max}$ (Eq. \ref{tildechi2}).  A well defined
      distribution is only present for the parameters that contribute
      to the PDF of Eq.~\ref{chi2ij}.  Parameters without any
      contribution lead to a uniform distribution.  }
\end{figure}

One of the significant advantages of AMIAS is that it determines
unambiguously the parameters to which the data are sensitive on i.e.
it determines $n_{\rm max}$ in the truncation of the infinite sum in
Eq.~(\ref{tildechi2}).  The results obtained are then invariant under
changes of $n_{\rm max}$.  The strategy is to increase $n_{\rm max}$
until there is no sensitivity to the additional exponentials and thus
no observable change in the sampled spectrum. In Fig.~\ref{nparams} we
show an example of such an analysis for the nucleon correlator $C(t)$,
which will be defined below.  As we increase $n_{\rm max}$, additional
exponential terms are identified resulting in a well defined
Gaussian-like distribution for each of the $E_n$ parameters. We can
get values for the parameters, $E_0$, $E_1$ and $E_2$, by fitting the
distribution of each mass to a Gaussian. For exponential terms beyond
the first three (e.g. those having exponents $E_3$ and $E_4$) we get a
uniform distribution indicating that there is no contribution to the
minimization of $\tilde{\chi}^2$ from these terms. We can thus safely
conclude that $n_{\rm max}=4$ is a safe choice since it includes, in
addition to the well determined exponentials, an additional {\it
  insensitive} exponential term. Table~\ref{nparams2} gives the
numerical values of the energy eigenstates corresponding to the
distributions of Fig.~\ref{nparams}, where we also compare with a
standard least squares minimization algorithm.  The advantage of AMIAS
is in the consistency of the results even when exponential terms are
included on which the data are {\it insensitive}.  With the usual
least squares minimization one can identify three states with an error
of over 50\% for the $E_2$. No indication as to the presence of higher
states can be extracted. AMIAS on the other hand yields $E_2$ with a
10\% error. As we increase the number of exponentials from 3 to 5 the
results for the parameters on which the correlator has sensitivity
remain unchanged demonstrating the robustness of the method.

\begin{table*}[ht]
  \centering
  \begin{tabular*}{1.\textwidth}{@{\extracolsep{\fill}}ccccccc}
    \hline\hline
     & \multicolumn{3}{c}{AMIAS}  &  \multicolumn{3}{c}{Standard Least Squares} \\
    $n_{\rm max}$ & $E_0$ & $E_1$ & $E_2$ & $E_0$ & $E_1$ & $E_2$ \\
    \hline\hline
    2   & $1.161(13)$ & $2.8990(32)$  & --           & $1.162(11)$  & $2.92(132)$ & --  \\
    3   & $1.1430(32)$ & $2.0453(98)$ & $5.0220(53)$ & $1.1439(23)$  & $2.03(89)$ & $5.022(2.7)$ \\
    4   & $1.1430(32)$ & $2.050(11)$  & $4.8850(64)$ & --            & --         & -- \\
    5   & $1.1432(32)$ & $2.052(11)$  & $4.8394(64)$ & --            & --         & -- \\
    \hline\hline
    \vspace*{0.1cm}
  \end{tabular*}
  \caption{The values obtained for the ground and first two excited states of $C^{(S,S)}_{11}$ from the B55.32 ensemble. These 
      values were obtained by a fitting a normal distribution to the results of Fig. \ref{nparams}. The units are in GeV. }
  \label{nparams2}
\end{table*}

\subsection{Fit range} 
\label{sec:fit_range}

\begin{figure*}[ht]
\medskip
\begin{minipage}{3.4in} 
\centering
\includegraphics[width=3.4in]{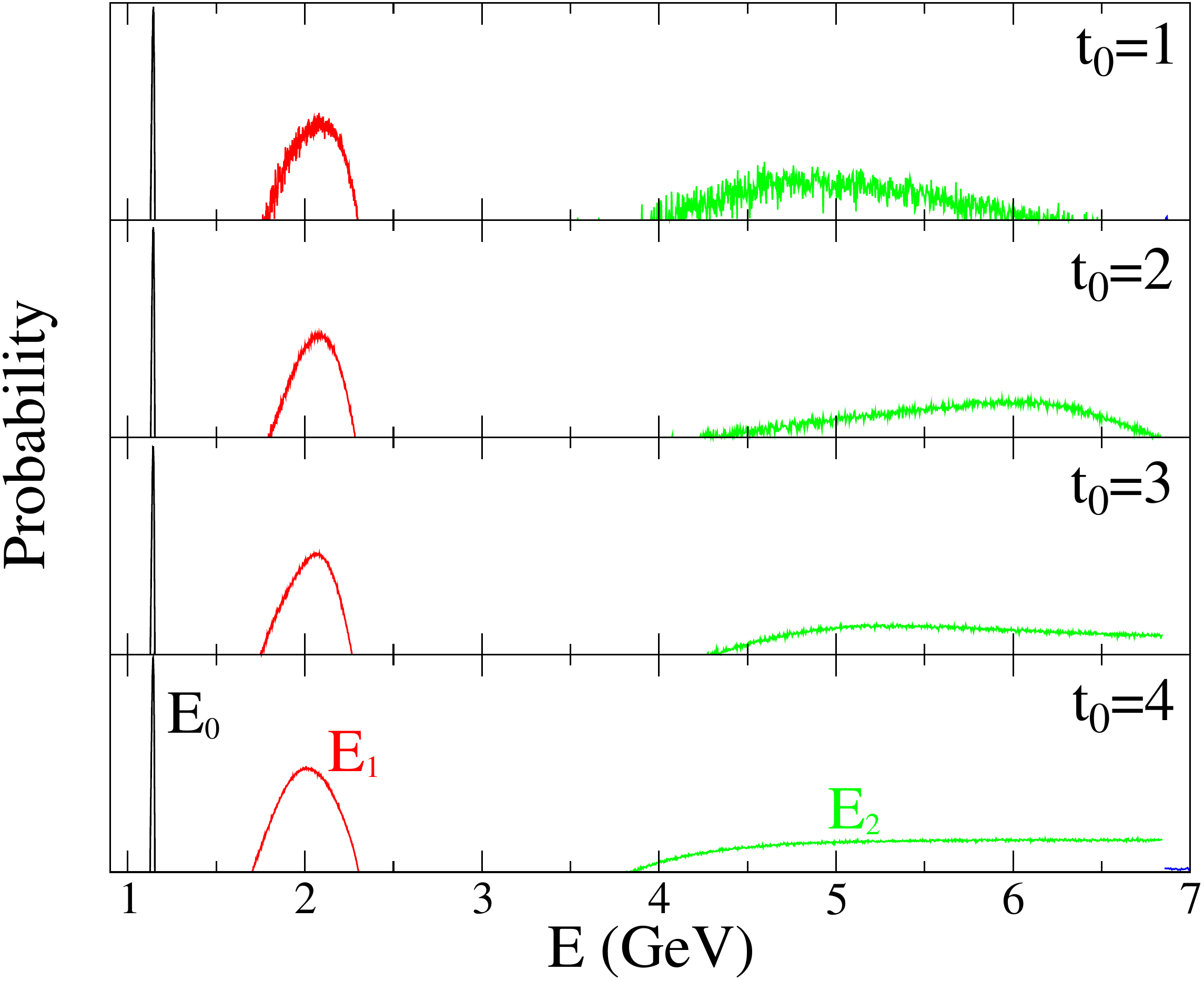}
\caption{\label{t0} The PDFs of the energies for an
  eight parameter case (four amplitudes and four energies). The
  horizontal axis is the energy in lattice units and the vertical axis is the sampled probability of Eq.~(\ref{critical}).  
  The fitting range is varied by changing $t_0$. Exponentials with larger exponents (excited states)
  are suppressed but the one with smaller exponents (low-lying spectrum) remains unaffected.
\label{t0_m}
}
\end{minipage}
\hspace{0.2cm}
\begin{minipage}{3.4in} 
\vspace{-0.2in}
\centering
\includegraphics[width=3.4in, keepaspectratio]{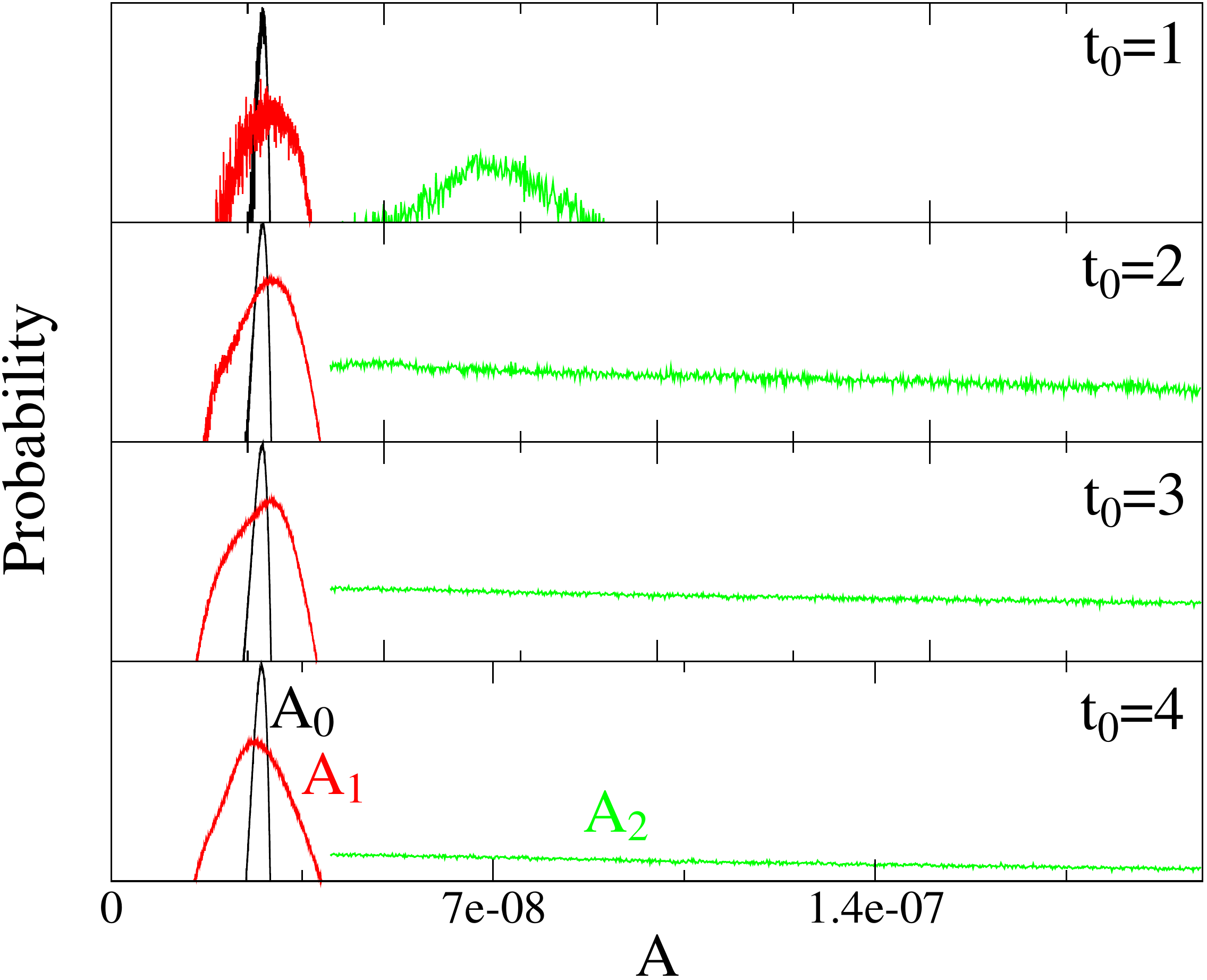}
\caption{
The PDFs of the amplitudes for an
  eight parameter case (four amplitudes and four masses). The
  horizontal axis is the
  amplitudes and the vertical axis is the sampled probability of Eq.~(\ref{critical}).  
  The fitting range is varied by changing $t_0$. 
  The suppressed excited state amplitudes appear with a uniform distribution.
\label{t0_A}}
\end{minipage}
\end{figure*}

Another parameter that enters in the analysis of correlators is the
initial time $t_0$ used in the fits to the exponential form. The
larger the initial time the smaller is the contribution of
exponentials beyond the first one.  Thus, we can suppress the
contribution from the higher exponents by varying the starting value
$t_0$. In the standard analysis the time $t_0$ is varied until the
so-called effective energy defined by 
\be E_{\rm
  eff}(t)=\log\frac{C(t)}{C(t+1)}\stackrel{t\rightarrow
  \infty}{\longrightarrow} E 
\ee 
becomes a constant as a function of
$t$ (plateau region) yielding the lowest exponent $E_0$ which will be
identified as the ground state energy in our lattice QCD study.
Similarly, in the AMIAS analysis we need to verify that the
determination of the exponents of the first exponential terms will be
unaffected as we vary $t_0$. In Figs.~\ref{t0_m} and \ref{t0_A} we
show such an analysis for both $E_n$ and $A_n$ is performed.  By
increasing $t_0$ the three exponential with exponent $E_2$ is
eliminated but the previous two exponents remain unaffected.
Elimination of the third exponent is also apparent when examining the
distribution of the amplitudes where the distribution of $A_2$ becomes
uniform as $t_0$ is increased.  Therefore, by increasing $t_0$ one
removes unnecessary correlations from terms that we cannot clearly
obtained within the available statistics and in addition, one verifies
the validity of the sampled dominant exponentials.

\subsection{Correlations} 

\begin{figure*}[ht!]
\begin{minipage}{3in}
\includegraphics[width=2.7in, keepaspectratio]{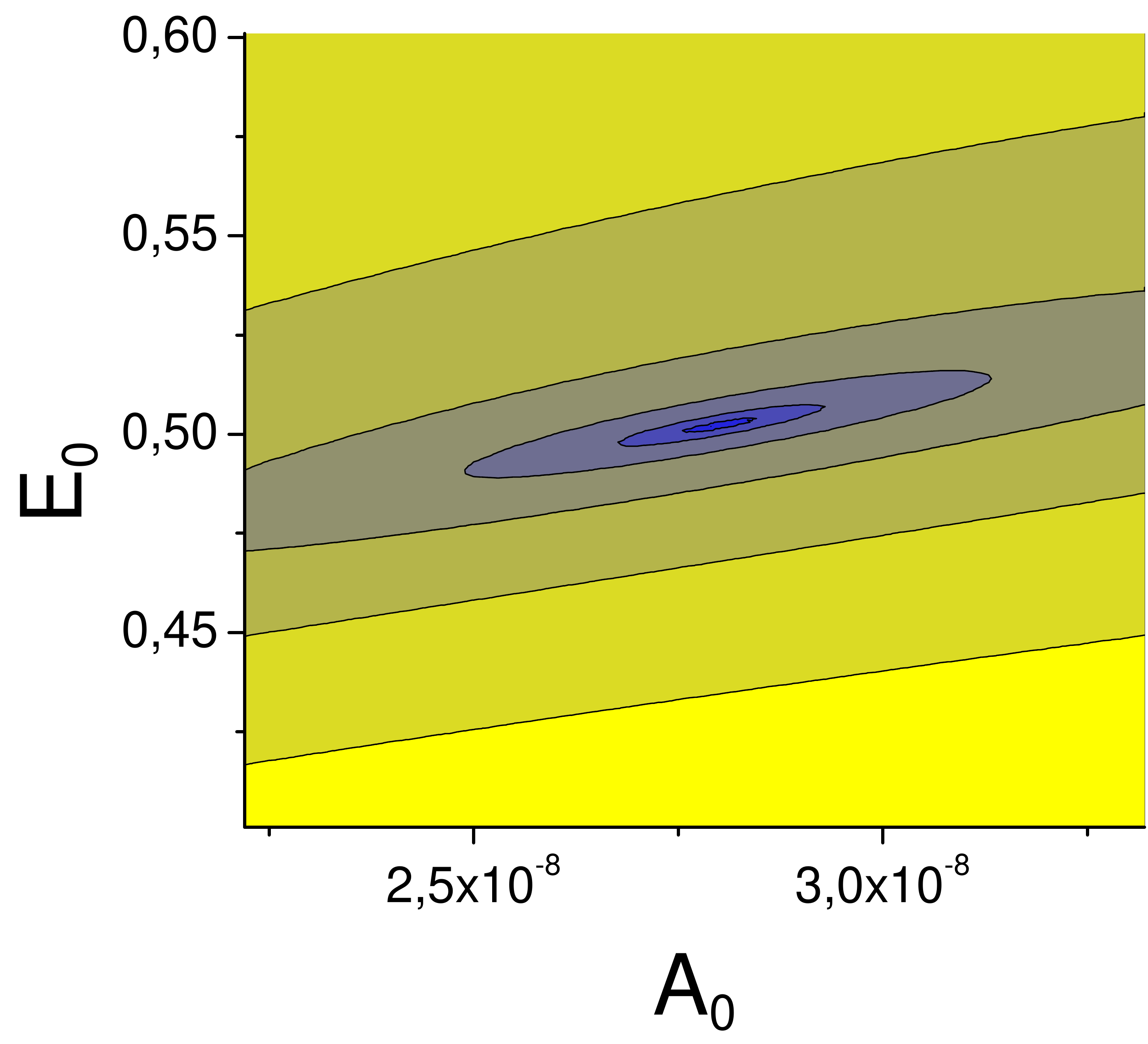}
\end{minipage}
\begin{minipage}{3in}
\includegraphics[width=2.7in, keepaspectratio]{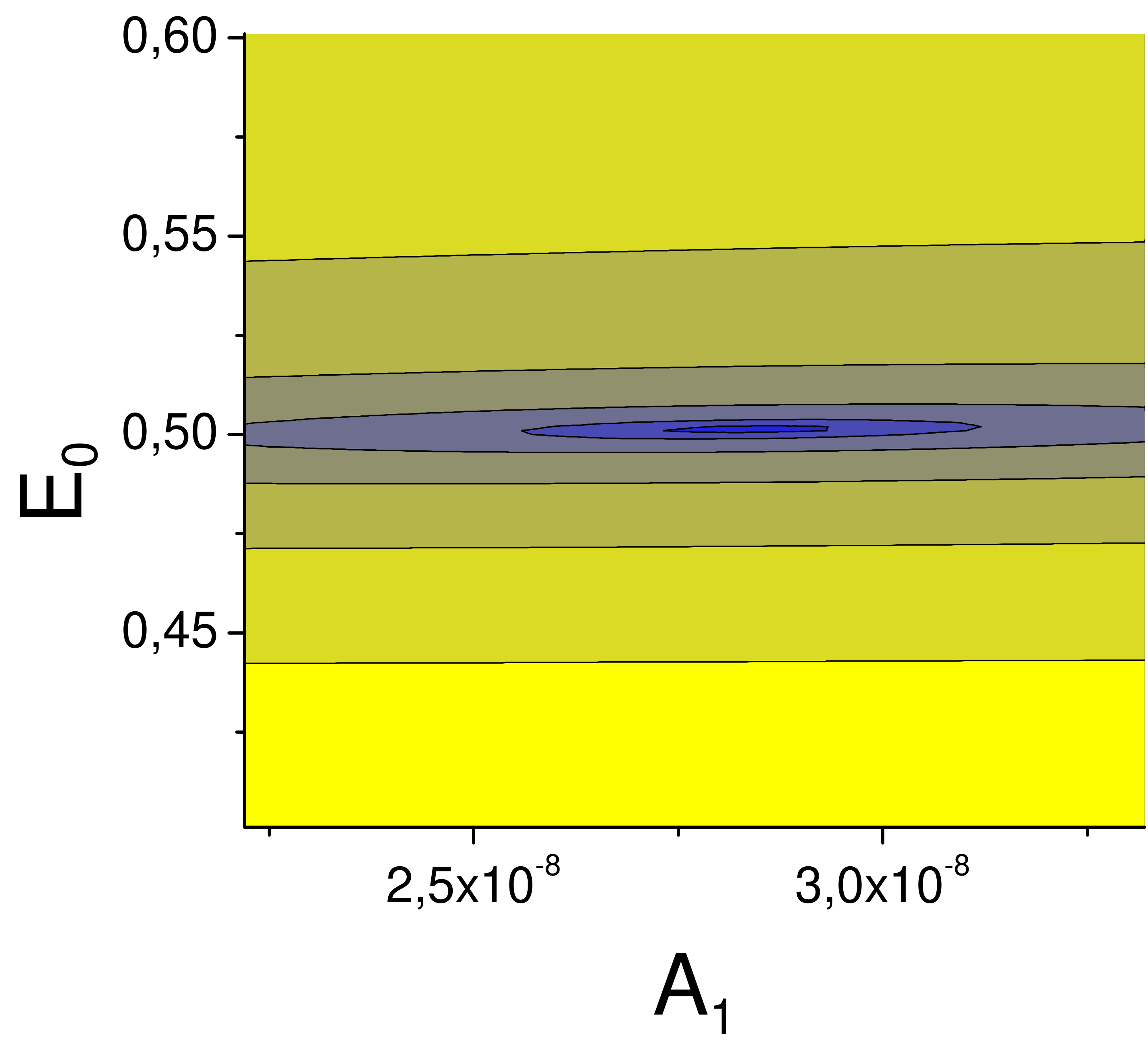}
\end{minipage}
\begin{minipage}{3in}
\includegraphics[width=2.7in, keepaspectratio]{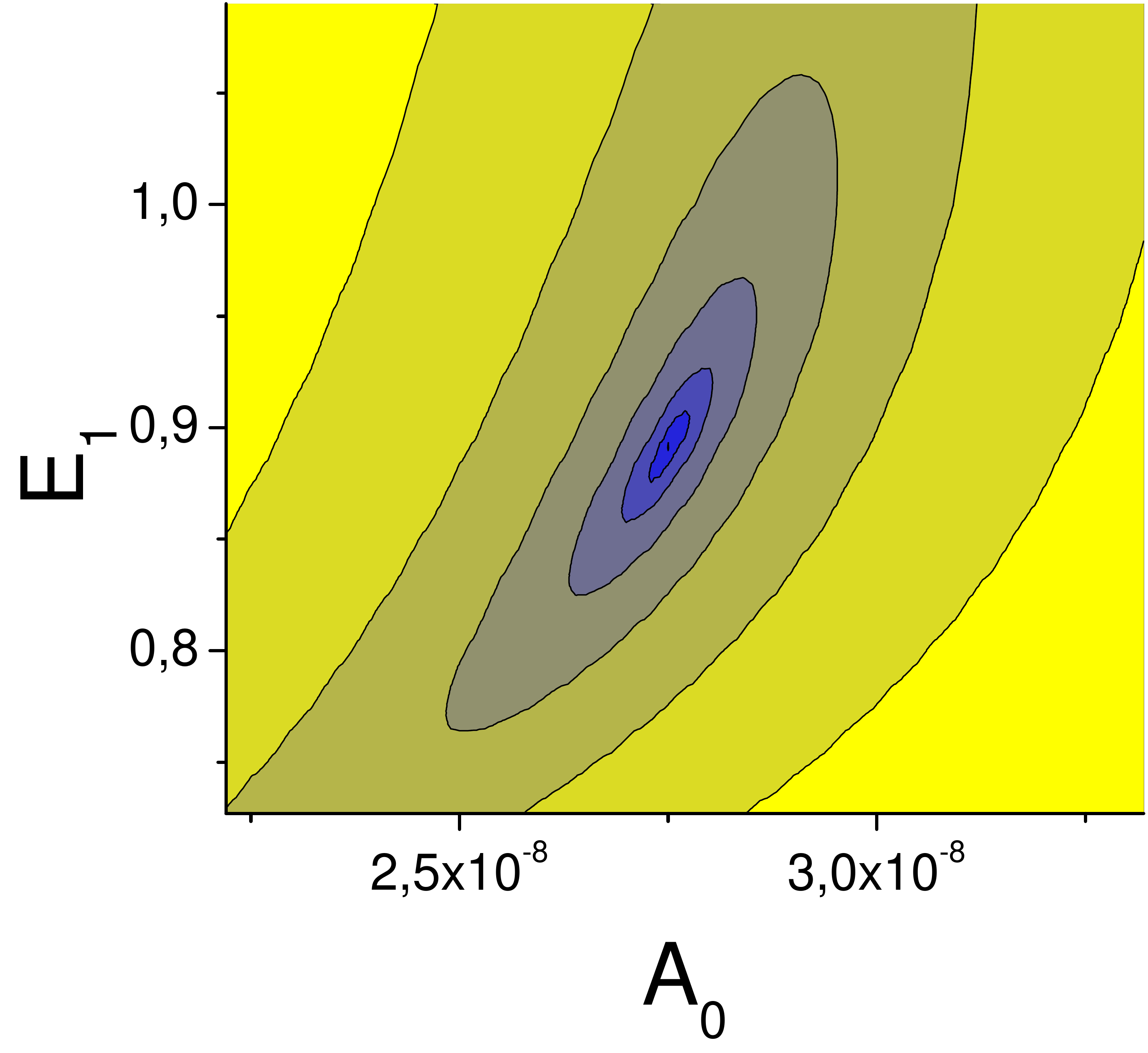}
\end{minipage}
\begin{minipage}{3in}
\includegraphics[width=2.7in, keepaspectratio]{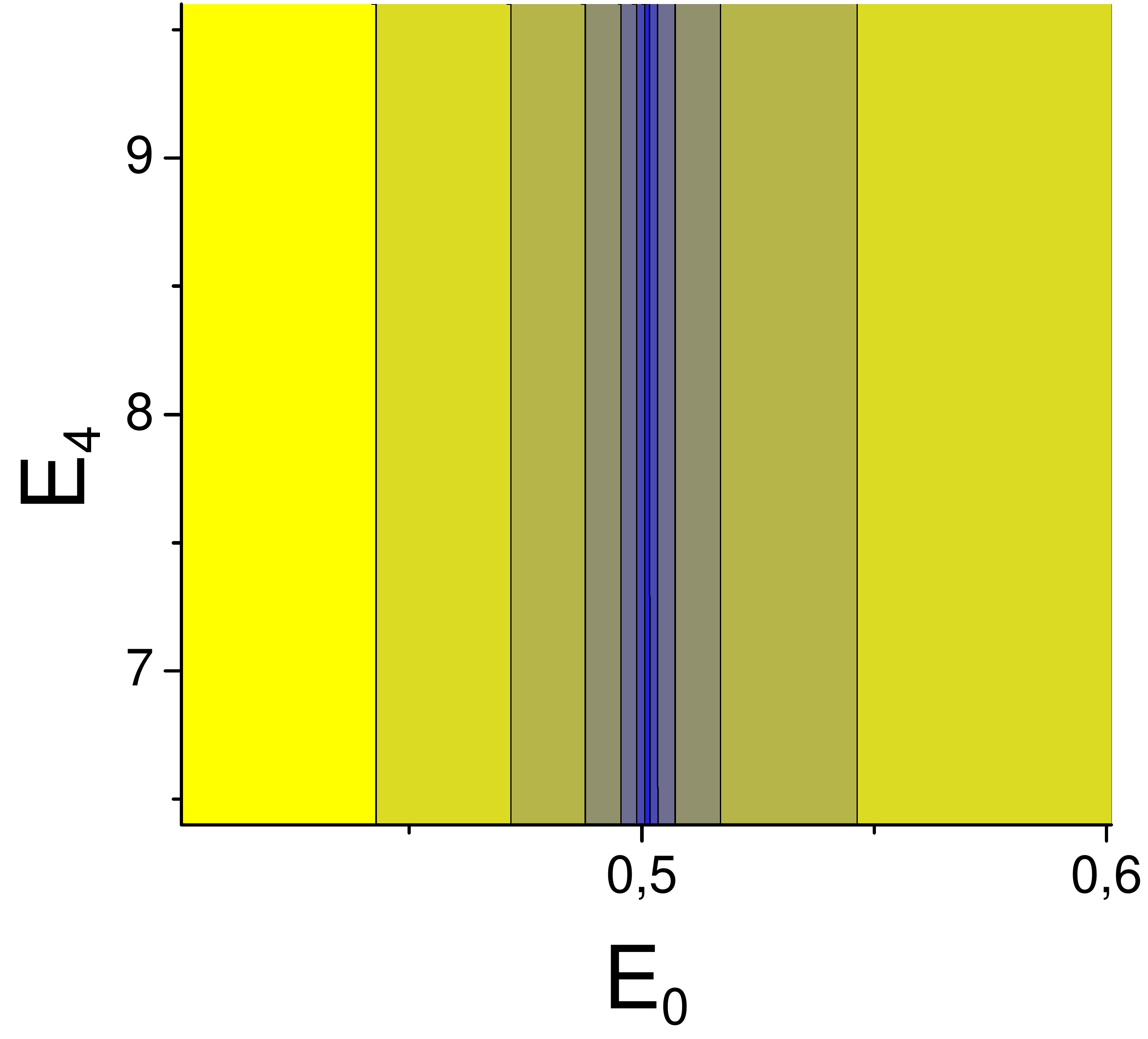}
\end{minipage}
\caption{Contour plots showing correlations among the parameters. 
 The contours
are color coded according to the $\chi^{2}$-value. 
The actual data used correspond to the nucleon correlator $C_{11}^{S,S}(t)$ of the B55.32 ensemble described in Section~\ref{sec:lattice_techniques}.
\label{correlations}}
\end{figure*}

A central issue that is properly treated in AMIAS, is the handling of
correlations, since all possible correlations are accounted for. The
sampling method of Eq.~(\ref{critical}) allows all fit parameters to
randomly vary and to yield solutions with all allowed values,
including the {\it insensitive} exponential terms.  The visualization of
the dominant correlations can be accomplished by a two-dimensional contour
plot in which the values of a selected pair of parameters is varied
around the region of maximum probability, keeping all other parameters
fixed. In Fig.~\ref{correlations} we present such a correlation analysis,
where the plane defined by the values of the parameters is color coded
according to the $\chi^{2}$-value. The top-left and bottom-left parts
are examples where the parameters are correlated, while the right part
is an example of uncorrelated parameters. In particular the
bottom-right part is an example where sampling is insensitive to one
of the parameters, namely $E_4$, indicating in this case the absence
of a fourth state (compared with Fig.~\ref{nparams}).

\section{Lattice techniques}
\label{sec:lattice_techniques}
In this section we give a more detailed description of the lattice QCD results
  that will be analyzed using AMIAS.

\subsection{Simulation details }

We use gauge configurations produced using the twisted mass fermion
action with two degenerate light quarks ($N_f=2$). We also use twisted
mass fermion gauge configurations adding to the light quarks a strange
and charm quark with masses fixed to their physical values
($N_f=2+1+1$). More details on the $N_f=2+1+1$ gauge configurations
produced by the European Twisted Mass Collaboration (ETMC) can be
found in Ref.~\cite{Baron:2010bv} and for the $N_f=2$ in
Ref.~\cite{Boucaud:2008xu}.  The twisted mass fermion ensembles used
in this work are summarized in Table~\ref{Table:params}.  The pion
mass values span a range from about 210~MeV to 450~MeV.  Apart from
the twisted mass fermion ensembles we also analyze an ensemble of $N_f
= 2$ clover fermion gauge configurations produced by the QCDSF
collaboration with near-physical pion mass of $m_\pi\simeq
160$~MeV~\cite{Bali:2012qs}. The nucleon excited states have been
analyzed on these gauge configurations using the conventional approach
in Ref.~\cite{Alexandrou:2013fsu} and thus serve as a good set for
making detailed comparison with the results obtained with AMIAS.

\begin{table}[h]
\begin{center}
\begin{tabular}{c|lllll}
\hline\hline
\multicolumn{5}{c}{$N_f=2$ ensembles}\\\hline
\multicolumn{5}{c}{ Twisted mass fermions, $a=0.0855(6)$~fm.}\\
$32^3\times 64$, & $m_\pi$~(GeV) & 0.2696(9)  & 0.3082(6)            \\
$L=2.74$~fm      & No. of confs  & 659        & 232       & &            \\\hline 
\multicolumn{5}{c}{ clover fermions, $a=0.0728(5)(19)$~fm.}\\
$48^3\times 64$, & $m_\pi$~(GeV) & 0.160 & &\\  
$L=2.8$~fm    & No. of confs  & 250        &  &   \\ \hline                 
\multicolumn{5}{c}{$N_f=2+1+1$ ensembles, $a=0.0863$~fm }\\\hline
$32^3\times 64$, & $m_\pi$~(GeV) & 0.375162           \\
(B55.32)      & No. of confs  & 5000        &        & &            \\\hline 
\end{tabular}
\caption{The  ensembles used in our analysis.}
\label{Table:params}
\end{center}
\vspace*{-.0cm}
\end{table}

\subsection{Correlator functions}

In order to study the energy spectrum within the framework of lattice QCD
one evaluates the Euclidean 
two-point correlation function $C(t)$ 
\bea
\nonumber
C({\bf p},t)&=&\sum_{\bf x} e^{i{\bf p}.{\bf x}} <J({\bf x},t) J^\dagger({\bf 0},0)>\\&=&\sum_{n=0}^\infty A_n e^{-E_n(p) t}\quad.
\label{two-point}
\eea where $J^\dagger({\bf x},t)$ is a creation operator acting on the QCD vacuum
(interpolating field) having the same
quantum numbers as the states of interest and ${\bf p}$ is the
three-momentum.  The two-point correlation function can be
expressed as a sum of the energy eigenstates of QCD that exponentially
decay as a function of time with an exponent that depends on the
energy of the state. Thus  by fitting to such a form one can in principle
extract the
energies $E_n$ of the system.  The problem, however, is that the higher
excited states decrease exponentially as compared to the ground state
and extracting them is difficult since the signal to noise decreases
exponentially  for all hadrons
except the lowest mass pseudoscalar. Fitting such correlation functions to two-exponentials
can be one approach to extract the first excited state when one has lattice data with small enough errors. On the other hand, extracting the
ground state is much easier since asymptotically it is the only state
that survives the large time limit of $C({\bf p},t)$.

Let us consider the zero momentum correlator $C(t)$ obtained from Eq.~(\ref{correlator0}) by setting ${\bf p}={\bf 0}$. In a simulation we  have
$N$  measurements at each time $t$,  with the $k^{\rm th}$ measurement having the
 form
$\{C^k(t_1),C^k(t_2),...,C^k(t_{N_t})\}$, where $N_t$ is the number
of lattice time slides on which the correlator is evaluated. Note that from now on we drop the discrete index on $t$.
The $N$ measurements are over a representative ensemble of gauge configurations
generated via a  Monte-Carlo sampling of the probability density
function of the Euclidean lattice QCD action. For each value of $t$ we
can form the average $C(t)=\frac{1}{N}\sum_{k=1}^{N}C^k(t)$, which was
the example used in the previous section to illustrate AMIAS. 

The results shown in Figs.~\ref{t0_m}, \ref{t0_A} and \ref{correlations} 
are carried out using the B55.32 ensemble of  $N_f=2+1+1$ twisted mass
 fermions of a  pion mass $m_\pi=373$~MeV,
 for which a large number of gauge configurations are generated. 
We will use this ensemble to perform  a detailed comparison of  AMIAS with the standard analysis used in lattice QCD
for the study  of excited states. 

The standard method to study excited states in lattice QCD is the
variational approach. One expands the basis to $N_{\rm inter}$ interpolating fields of the quantum numbers of states of interest and
construct a correlation matrix. In this work we consider a correlation matrix of the form
\bea
\label{correlation_matrix}
\begin{split}
&C^{\pm,(i,j)}_{ab}({\bf p},t)=\sum_{\bf x}e^{i{\bf p}.{\bf x}}\textrm{Tr}[\frac{1}{4}(1\pm\gamma_0)\langle J_a^{(i)}({\bf x}, t)\bar{J}_b^{(j)}({\bf 0},0)\rangle] \nonumber\\
&= \sum_{n=0}^\infty e^{-E_nt }\textrm{Tr}[\frac{1}{4}(1\pm\gamma_0)\langle 0|J_a^{(i)}|n\rangle \langle n|J_b^{(j)}|0\rangle]\,,&\nonumber\\
&\begin{array}{l}
  i,j=1,\ldots, {N_G}\\a,b=1,2,
\end{array}
\end{split}
\eea
where we denote the different types of interpolating fields
with two type of indices  $i,j$ and $a,b$.  
The superscripts $i,j$ on
the correlation matrix $C^{\pm}(t)$ correspond to different levels of
Gaussian smearing, while  the subscripts $a,b$ to different spin combinations of
the nucleon interpolating fields given by:

\bea
J_1^{(i)}=(u^{\rm T\, i}C\gamma_5d^i)u^i\,\,\, {\rm and}\,\,\,J_2^{(i)}=(u^{\rm T\,i}Cd^i)\gamma_5u^i. \label{corrs}
\eea
Gaussian smearing is applied to the quark fields using the hopping matrix 
$H(\vec x,\vec y; U(t))$:
\bea
u^{(a,i)}(t,\vec x) &=& \sum_{\vec y} F^{ab}(\vec x,\vec y;U(t))\ u^b(t,\vec y)\,,\\
F &=& (    \mathbb{1} + {a_G} H)^{N_G} \,, \nonumber\\
H(\vec x,\vec y; U(t)) &=& \sum_{i=1}^3[U_i(x) \delta_{x,y-\hat\imath} + U_i^\dagger(x-\hat\imath) \delta_{x,y+\hat\imath}]\,. \nonumber
\eea
We also apply APE-smearing to the gauge fields $U_\mu$ entering
 $H$. The parameters for the Gaussian smearing
$a_G$ and $N_G$ are optimized using the nucleon ground state~\cite{Alexandrou:2013jsa}.
The local nucleon interpolator, $J_1^{(i)}$, is well
known to have a good overlap with the ground state of the nucleon.
The trace in Eq.~(\ref{correlation_matrix}) is taken over Dirac indices and 
the correlation matrices $C^+(t)$ and $C^-(t)$
yield the positive and negative parity states of the nucleon, respectively.
 The states
$|n\rangle$ are eigenstates of the Hamiltonian with $E_n< E_{n+1}$  and
we assume that the temporal extent of the lattice is large
enough to neglect boundary contributions. 

\begin{figure}[ht!]
\medskip
\centering
\includegraphics[width=3in, keepaspectratio]{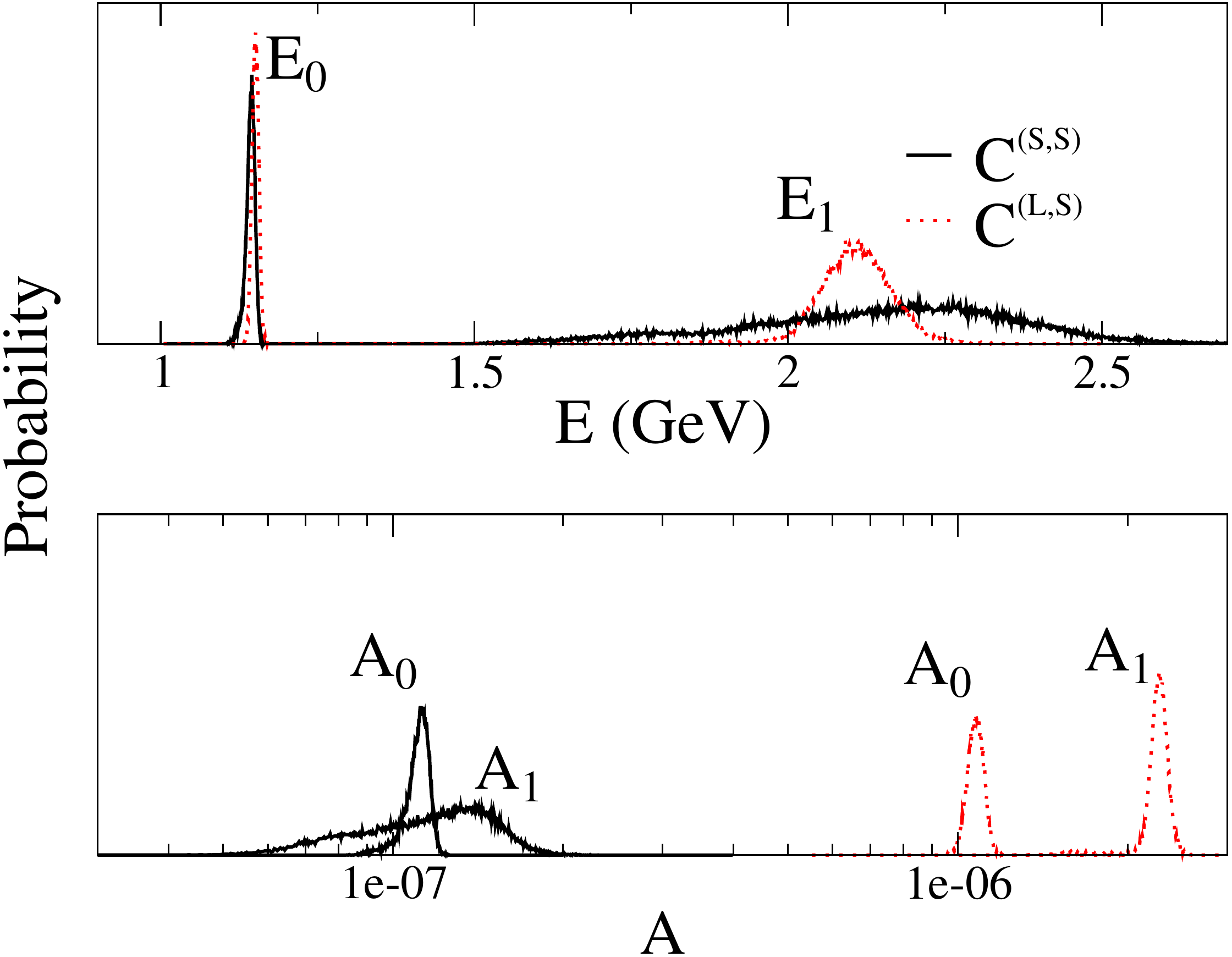}
\caption{The PDFs for the ground and excited energies (top) and amplitudes (bottom)
  for smeared-smeared and local-smeared, correlators using the B55.32
  ensemble.  As expected the amplitudes depend on the smearing level
  while the energies are statistically equivalent (a reduction in the
  level of smearing leads to more well defined distributions for the
  excited states).
\label{localsmeared}}
\end{figure}

The low-lying energy spectrum   should be unaffected by
what nucleon interpolating field is used and in particular how much
smearing is applied as determined by the parameters $N_G$ and
$\alpha_G$. This is because Gaussian smearing only effects the
coupling of the interpolating field with the nucleon eigenstates
$|n\rangle$ i.e. only the $A_n^{i}$ are affected but not the $E_n^i$. As long as the amplitudes are non-zero
the energy of interest can be extracted. 
In Fig.~\ref{localsmeared} we verify the independency of the energies
by comparing the results from a smeared-smeared correlator $C^{(S,S)}$ where both
interpolating fields in Eq.~(\ref{correlator}) are smeared to results
extracted from a local-smeared correlator $C^{(L,S)}$, where only the interpolating
field at the source  is smeared. We show results for the ground and first excited
states. As expected the
distributions for $E_0$ and $E_1$ are statistically equivalent but the
local-smeared correlator provides a better estimate for the excited
state since it has less overlap with the ground state. In contrast,
the amplitudes $A_0$ and $A_1$ depend on the level of smearing. This
is an indication that reducing smearing will improve the results
extracted for the excited states, since their contribution to the
correlator will be larger.

\subsection{Comparison of AMIAS with the variational method}

The standard way to extract the ground state energy in lattice QCD is
to probe the long time-limit of the correlation function and
consider the effective energy  $E_{\rm eff}(t)$, which
becomes independent of $t$ when the ground state is the dominant
contribution to the correlator. Fitting the effective
energy (or mass) to a constant in the plateau region yields the ground state energy (or lowest mass).
As has already been demonstrated in Ref.~\cite{Alexandrou:2008bp}, the value
obtained with AMIAS for the ground state is in agreement with the
one extracted from the effective mass plateaus. 
In this section, we focus our attention to the extraction of the
excited states.

In order to study excited states in lattice QCD, one usually 
applies the variational method and defines a 
generalized eigenvalue problem (GEVP) given by
\be \label{GEVP}
\begin{split}
&C(t)v_n(t,t_i)=\lambda_n(t,t_i)C(t_i)v_n(t,t_i),\hspace*{0.5cm} \\
&n=1,\ldots, N, \, t>t_i\,,
\end{split}
\ee
where $E_n=\lim_{t\rightarrow \infty} -\partial_t \log
\lambda_n(t,t_i)$. The corrections to $E_n$ decrease exponentially
like $e^{-\Delta E_n t}$ where $\Delta E_n=\min_{m\neq
  n}|E_m-E_n|$~\cite{Luscher:1990ck} for fixed $t_i$. In a recent
application of the variational method for the analysis of the
nucleon states~\cite{Alexandrou:2013fsu}
 we found that  a variational basis constructed from different
smearing levels of the standard interpolating field $J_1$ 
 containing both a small and a large number of
Gaussian smearings  $N_G$ is an
 appropriate basis  for extracting the first positive parity excited state, known as the Roper.  

\begin{figure}[h!]
\medskip
\includegraphics[width=3.4in, keepaspectratio]{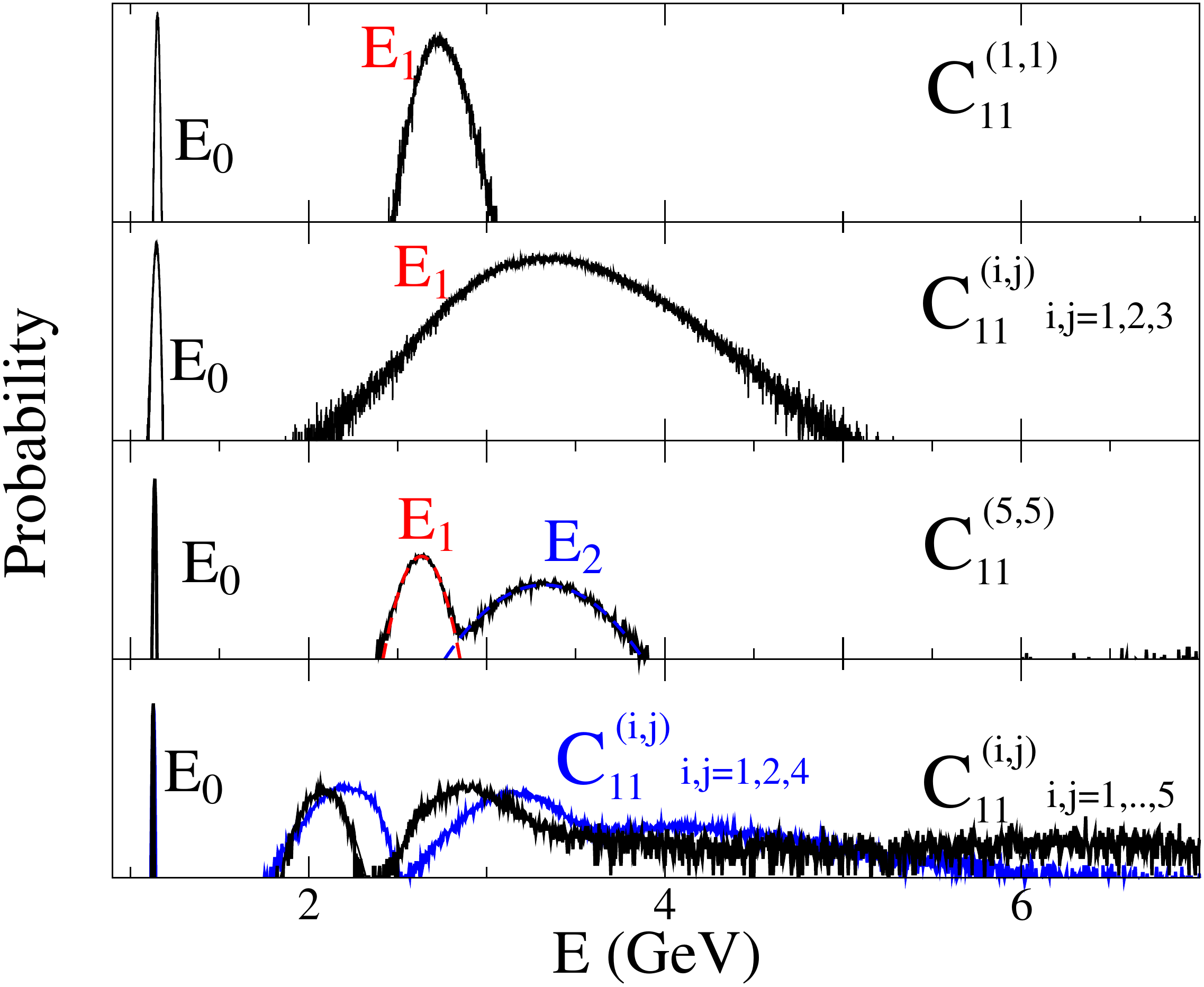}
\caption{\label{amias5} The spectrum from AMIAS when using the
  diagonal $C_{11}^{(1,1)}$ and $C_{11}^{(5,5)}$ correlators as well
  as the $3\times 3$ correlation matrices $C_{1,1}^{(i,j)},\,
  i,j=1,2,3$ and $C_{1,1}^{(i,j)},\, i,j=1,2,4$. Results are also
  shown for the $5\times 5$ matrix $C_{1,1}^{(i,j)},\, i,j=1,\ldots,
  5$. The inclusion of a variational basis that includes small and large
  smearings leads to an improved excited-state spectrum.
}
\end{figure}

\begin{figure*}[ht!]
\medskip
\centering
\includegraphics[width=7in, keepaspectratio]{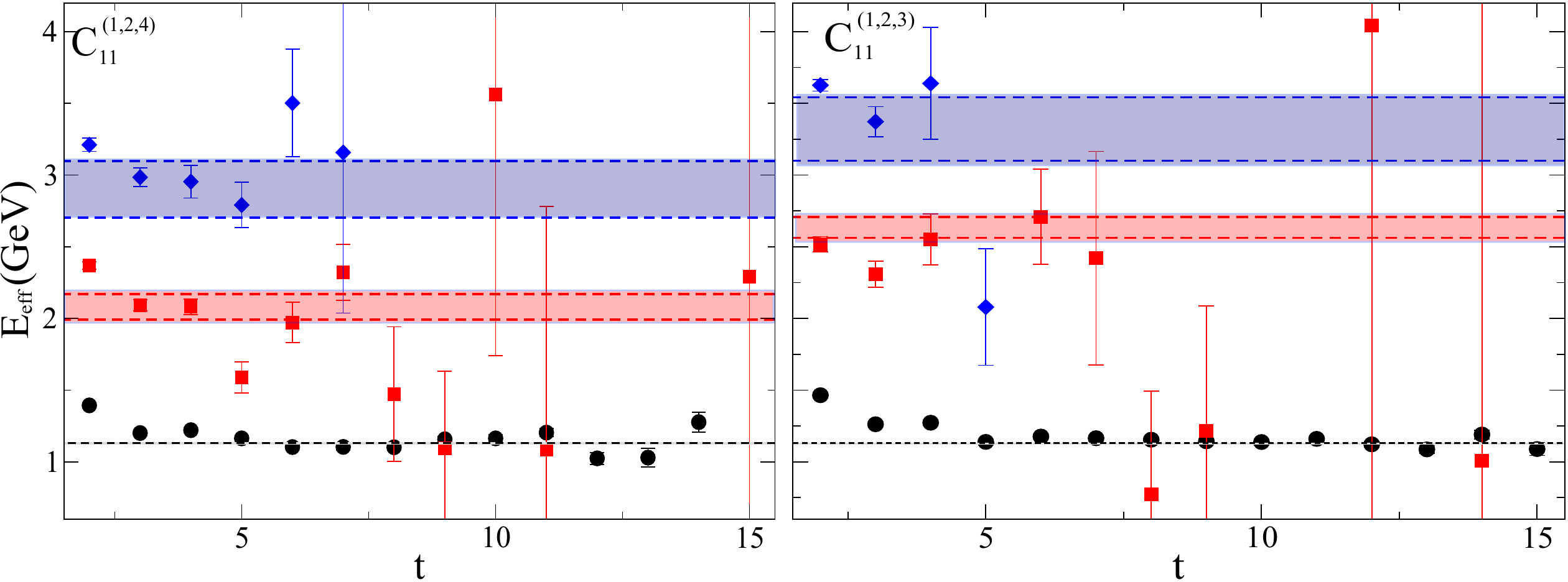}
\caption{\label{amiasvsgevp} The points show the effective masses determining 
from the GEVP analysis while the bands show the values obtained by AMIAS for the same correlation 
matrix. The left panel shows results extracted  the $3\times 3$ correlation matrix $C_{11}^{(1,2,4)}$, while the left panel 
from the $3\times 3$ correlation matrix, $C_{11}^{(1,2,3)}$.
}
\end{figure*}

In order to compare AMIAS with the results extracted using GEVP, we
consider the same variational basis used in that study, namely we set
the values of the smearing parameters to $\alpha_G=4.0$ and $N_G=$10,
30, 50, 180 and 300.  These different smearing levels are labeled by
the superscript $i=1,\ldots, 5$ on $J_a^{(i)}$ and produce a source
with a root mean square radius $2.26a,~3.77a,~4.77a,~8.37a$ and
$10.19a$ in units of the lattice spacing $a$, respectively.  The
resulting matrices are symmetrized. For this analysis, we use 200
twisted mass configurations produced at $\beta$=3.9, a$\mu$ = 0.004 or
m$_\pi\sim 308$~MeV on a 32$^3$ $\times$ 64 lattice.  In
Fig.~\ref{amias5} we show the results extracted from AMIAS for the
various correlation matrices. The value for the first excited state
obtained from fitting the diagonal elements is statistically
equivalent for all values of the smearings. This is illustrated in
Fig.~\ref{amias5}, where we show the results extracted from
$C_{1,1}^{(1,1)}$ and $C_{1,1}^{(5,5)}$, which correspond to the
smallest and largest smearings, respectively. When a correlation
matrix is used there is a lowering in the value of the first excited
state and in addition a second excited state appears close-by. This
indicates that this state cannot be detected in the diagonal
correlators, presumably having a very small overlap within the
standard nucleon interpolating field. A $3\times 3$ correlation matrix
using the first three smearing levels $C_{1,1}^{(i,j)}, \,i,j=1,2,3$
is compared with a $3\times 3$ matrix, $C_{1,1}^{(i,j)},\,i,j=1,2,4$.
Using a variational basis that includes small and large smearings i.e
$N_G=10$ and $N_G=300$ results in a lowering of the energy of the
first excited state.  Furthermore, using the full $5\times 5$
correlation matrix, $C_{1,1}^{(i,j)},\, i,j=1,\ldots, 5$, results in a
reduction of the error but leaves the mean values unaffected.

We compare the results extracted with AMIAS to the effective energies
extracted from GEVP in Fig.~\ref{amiasvsgevp}. The AMIAS values are
shown by the bands, while the effective mass by the symbols. As can be
seen, the effective energies as a function of the time separation
become very noisy for the first and second excited states making the
identification of the plateau region ambiguous.  AMIAS nicely extracts
these energies, which are in agreement with the levels that the
variational approach seems to indicate.  This illustrates the
advantage of AMIAS in the determination of the excited states.

\section{The low-lying nucleon spectrum}
\label{sec:results}

Having presented a detailed comparison with the results obtained by
using GEVP, in this section we present the results for the nucleon
excited states in the positive- and negative-parity channels obtained
by applying AMIAS.  For this analysis we use the $4\times4$
correlation matrix $C_{ab}^{(i,j)},~~a,b=1,2~~i,j=3,4$. These
correlation matrices were also utilized within the GEVP analysis of
Ref.~\cite{Alexandrou:2013fsu} and thus readily provide a comparison
with the results obtained within the AMIAS analysis.

\begin{figure}[h!]
  \includegraphics[width=3.4in, keepaspectratio]{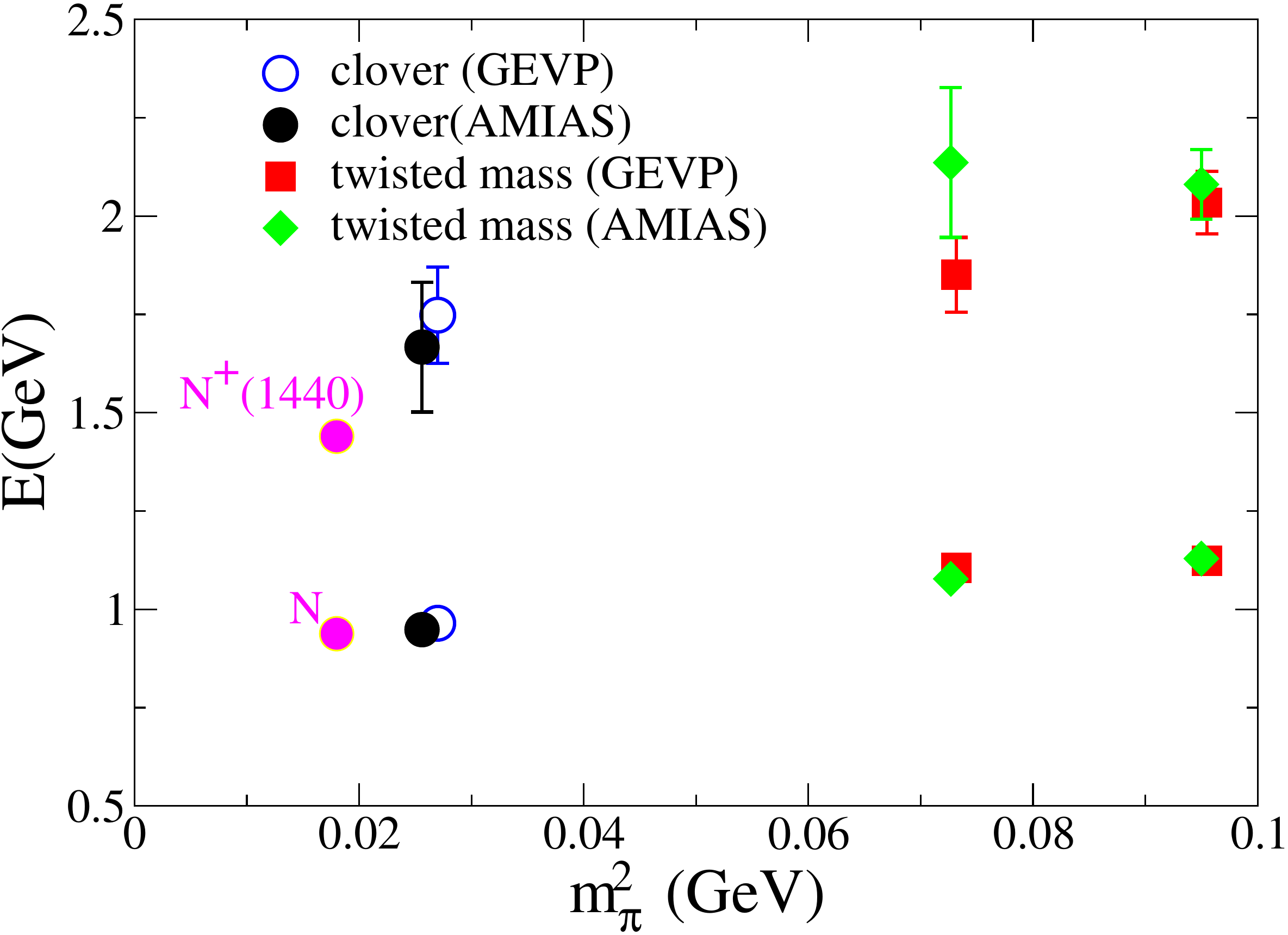}
  \caption{\label{results_pos} The nucleon ground state and first
    excited state in the positive parity channel.  The values obtained
    using AMIAS (filled back circles and filled green diamond) are
    compared to the results extracted from
    GEVP~\cite{Alexandrou:2013fsu}. }
\end{figure}
\begin{figure}[h!]
  \centering
  \includegraphics[width=4in, keepaspectratio]{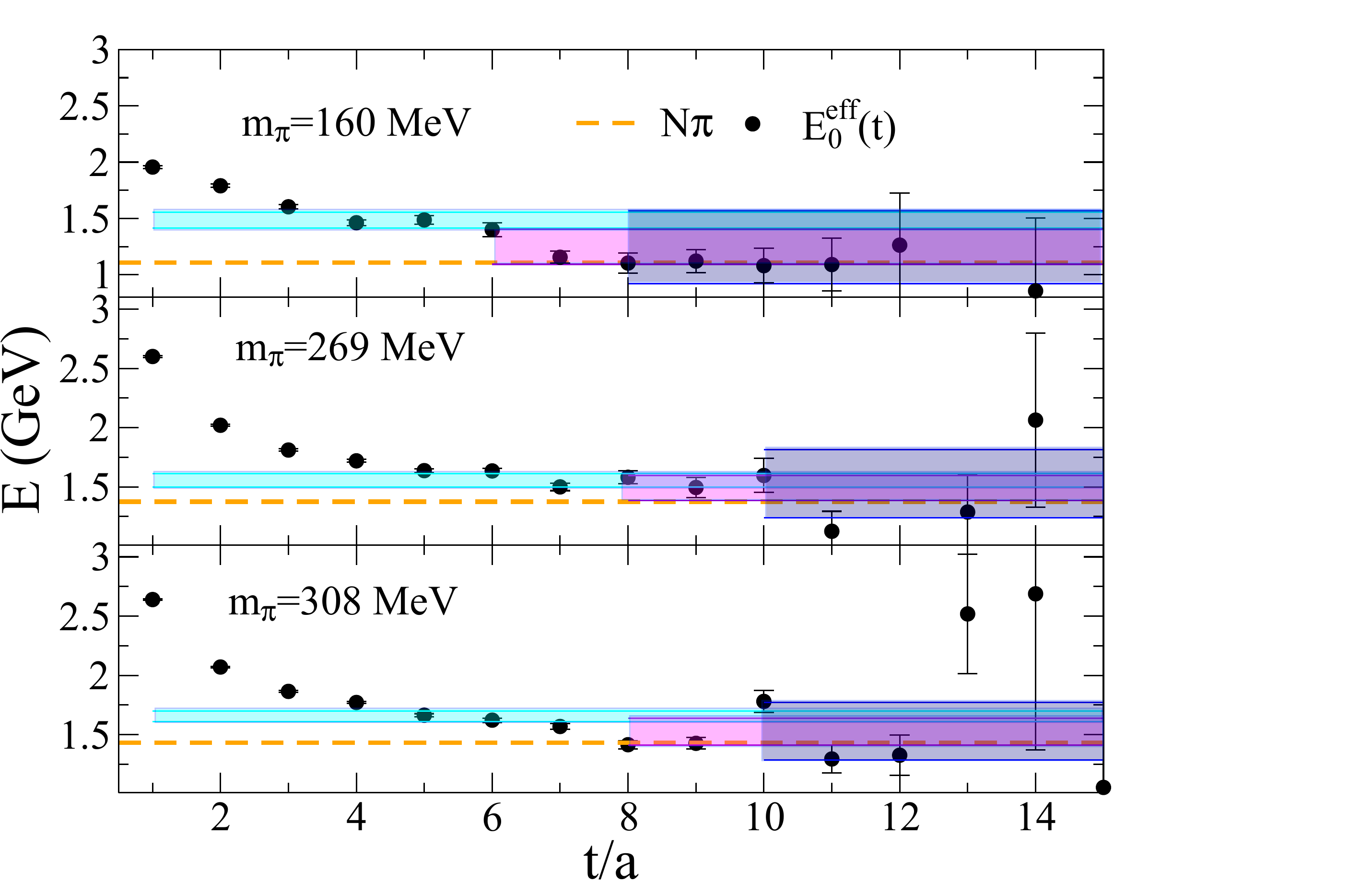}
  \caption{\label{negpar_piN} The effective mass for the negative
    parity ground state for all of our ensembles. Shaded error bands
    obtained by AMIAS for various values of $t_0$  are shown.
    The dotted line indicates the $N\pi$ ground state energy which is
    obtained from the sum of the nucleon and pion masses for the
    particular ensemble.}
\end{figure}

\begin{figure}[h!]
  \centering
  \includegraphics[width=3.4in, keepaspectratio]{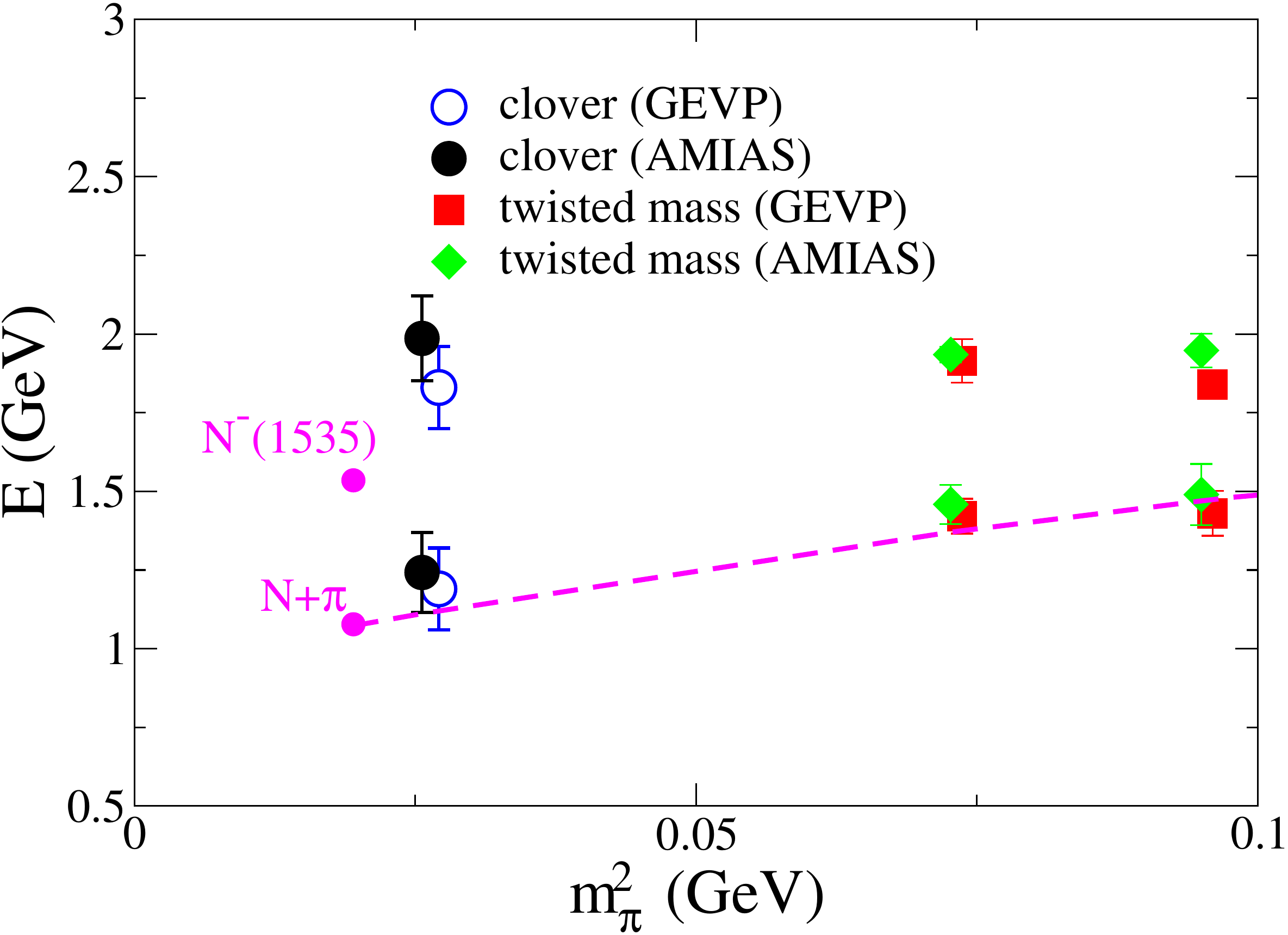}
  \caption{\label{results_neg} The same as in Fig.~\ref{results_pos} but for the negative parity channel.}
\end{figure}

In Figs.~\ref{results_pos} and \ref{results_neg} we show the results
for both positive and negative parity states, for the twisted mass
ensembles and for the clover ensemble analyzed in this work. The
results are compared with those obtained using the GEVP. For the case
of the clover action and the heaviest of the twisted mass ensembles
($m_\pi=0.3082(6)$GeV) we have refined our analysis by expanding the
basis of the GEVP as compared to the analysis carried out in
Ref.~\cite{Alexandrou:2013fsu}.  For these two ensembles the results
obtained with AMIAS are consistent with the GEVP. For the twisted mass
ensemble with the smaller pion ($m_\pi=0.2696(9)$GeV) the first
excited state obtained with AMIAS has a larger error than the one
extracted from the plateau.  This is because in the GEVP one needs to
fit for small $t_0$ in order to have enough fitting points. Contrary
in AMIAS we utilize the whole time dependence and thus we are more
confident for the AMIAS determination.

For the negative parity channel there is an interesting feature
connected to the $N\pi$ scattering state.  Unlike the behavior
observed in the positive channel where AMIAS yields results that did
not show any dependence on $t_0$ here this is no longer the case. The
lowest level extracted when using $t_0/a=1$ is higher that the energy
of the S-wave $N\pi$ scattering state assuming that no interactions take place.
AMIAS yields
results in agreement with the GEVP only after increasing the value of
$t_0$, albeit with larger errors. This was further investigated in
Fig.~\ref{negpar_piN}, where we show the effective mass for the
negative parity ground state for the three lowest pion masses used.
The values extracted using AMIAS for various starting values of the
fit range, $t_0$ are shown by the shaded error bands as $t_0$. The
values of $t_0$ used are indicated by the timeslice where the bands
start. The dashed line shows the energy of the scattering state being
the sum of the pion and nucleon mass.  As can be see, with increasing
$t_0$ the error band width increases becoming consistent with the mass
of the $N\pi$ scattering state but at the same remains consistent with
the band obtained with $t_0/a=1$. This is consistent with our analysis
in section~\ref{sec:fit_range} where we have shown that increasing
$t_0$ suppresses excited states but does not affect the mean value of
the ground state. One possible reason that one needs to increase $t_0$
to obtain agreement with the $N\pi$ energy maybe the small overlap
this two-particle state may have with a single particle interpolating
field, requiring the suppression of the excited states to get a signal
and may explain why other groups have not been able to detect it.  In
Fig. \ref{results_neg} we give the values extracted for the ground and
the first excited state in the negative channel. The values shown for
the $N\pi$ correspond to the magenta bands in Fig.~\ref{negpar_piN}
obtained using $t_0/a=6$ for clover fermions and $t_0/a=8$ for
twisted mass fermions. In contrast, the excited state was obtained
using $t_0/a=1$ in analogy with the positive parity channel.

\section{Summary and Conclusions}
\label{sec:conclusions}

A novel method for the analysis of excited states within lattice QCD
is applied to study the spectrum of the nucleon. The method uses
importance sampling to probe the correct minimum of the
multidimensional space defined by the states of the correlation
matrix. AMIAS determines the number of excited states, which can be
extracted from the information that the lattice data encode. It takes
advantage of the whole time dependence of the correlators not
requiring identification of the large time asymptot that limits the
accuracy of the determination of excited states.  The method is
applied to successfully extract the first excited states in the
positive and parity nucleon channels, identifying the Roper in the
positive parity channel.  In the negative parity it can extract the
energy of the first excited state using the entire time dependence of
the two-point function the energy of the $N\pi$ scattering state is
only extracted after eliminating the initial few time slices
suppressing the excited states.  This is a feature that we plan to
explore further in future studies.

\section*{Acknowledgments}
We thank G. Koutsou for providing the two point functions for the
clover fermion ensemble.
Numerical calculations have used the Cy-Tera facility of the Cyprus
Institute under the project Cy-Tera (NEA
Y$\Pi$O$\Delta$OMH/$\Sigma$TPATH/0308/31) funded by the Cyprus Research Promotion Foundation.

\bibliography{AMIAS_2012}
\end{document}